# Design, fabrication and measurement of a novel cooling arm for fusion energy source

JIANG Shui-Dong (江水东)[1,2]   LIU Jing-Quan (刘景全)[1,2]   MEI Jia-Bin (梅加兵)[1,2]   YANG Bin (杨斌)[1,2]
YANG Chun-Sheng (杨春生)[1,2]

[1]National Key Laboratory of Science and Technology on Nano / Micro Fabrication, Shanghai 100240, China

[2]Institute of Mico and Nano science and Technology, Shanghai Jiao Tong University, Shanghai 100240, China

JIANG Shui-Dong :jiangshuidong@sjtu.edu.cn

LIU Jing-Quan :jqliu@sjtu.edu.cn (corresponding author)

**Abstract**: The issues of energy and environment are the main constraint of sustainable development in worldwide. Nuclear energy source is one important optional choice for long term sustainable development. The nuclear energy consists of fusion energy and fission energy. Compared with fission, inertial confinement fusion (ICF) is a kind of clean fusion energy and can generate large energy and little environmental pollution. ICF mainly consists of peripheral driver unit and target. The cooling arm is an important component of the target, which cools the hohlraum to maintain the required temperature and positions the thermal-mechanical package (TMP) assembly. This paper mainly investigates the cooling arm, including the structural design, the verticality of sidewall and the mechanical properties. The TMP assembly is uniformly clamped in its radial when using (111) crystal orientation silicon to fabricate cooling arm. The finite element method is used to design the structure of cooling arm with 16 clamping arms, and the MEMS technologies are employed to fabricate the micro-size cooling arm structure with high vertical sidewall. Finally, the mechanical test of cooling arm is taken, and the result can meet the requirement of positioning TMP assembly.

## 1  Introduction

With high development of industry, more and more energy will be consumed, but the earth's energy resources are limited. With the traditional and non-renewable energy continuing to run out, we have an urgent requirement to find new energy to sustain the needs of human survival. Fusion, solar energy and biofuels are the energy sources satisfying the power need for the future without the negative environmental impacts. The simplest fusion fuels, the heavy isotopes of hydrogen (deuterium and tritium), can be derived from water, which is relatively abundant resource. The fuels are virtually inexhaustible, one in every 6,500 atoms on Earth is a deuterium atom, and they are available worldwide. One gallon of seawater would provide the equivalent energy of 300 gallons of gasoline; fuel from 50 cups of water contains the energy equivalent of two tons of coal. A fusion power plant would produce no climate-changing gases, as well as considerably lower amounts and less environmentally harmful radioactive byproducts than current nuclear power plants. And there would be no danger of a runaway reaction or core meltdown in a fusion power plant [1].

Inertial confinement fusion (ICF) is a process in which nuclear fusion reactions are initiated by heating and compressing a fuel capsule, typically a pellet that most often contains a mixture of deuterium and tritium. The powerful laser beams impinge on the inside of a hohlraum, where the laser energy is converted to X-ray energy. These X-rays bathe the fuel capsule and rapidly ablate the capsule's outer layer. The principle of conservation of momentum forces the remaining material to implode or compress. Compressing the deuterium and tritium (D-T) fuel that has been formed in an ice layer inside the capsule to extraordinarily high temperature, pressure and density ignites the burning hydrogen plasma. Thermonuclear burn spreads rapidly through the compressed fuel, yielding many times the input energy [2-5].

All of the ignition experiments require targets. Fig.1 is the schematic diagram of the cryogenic target [2]. To achieve ignition, the target contains four essential components: deuterium and tritium (D-T) fuel, a capsule with fill tube, a hohlraum and thermal control hardware. The geometric shape of the D-T ice is determined by the heat transfer between the hohlraum and the capsule. Precise control of the heat transfer plays an important role in forming a spherical D-T ice layer. This is achieved through the TMP assembly that consists of the hohlraum and aluminum cylinder. Two cooling arms, as a precise heat transfer part, are used to maintain the required temperature [2-4]. The other function cooling arm in this target is to position TMP assembly.

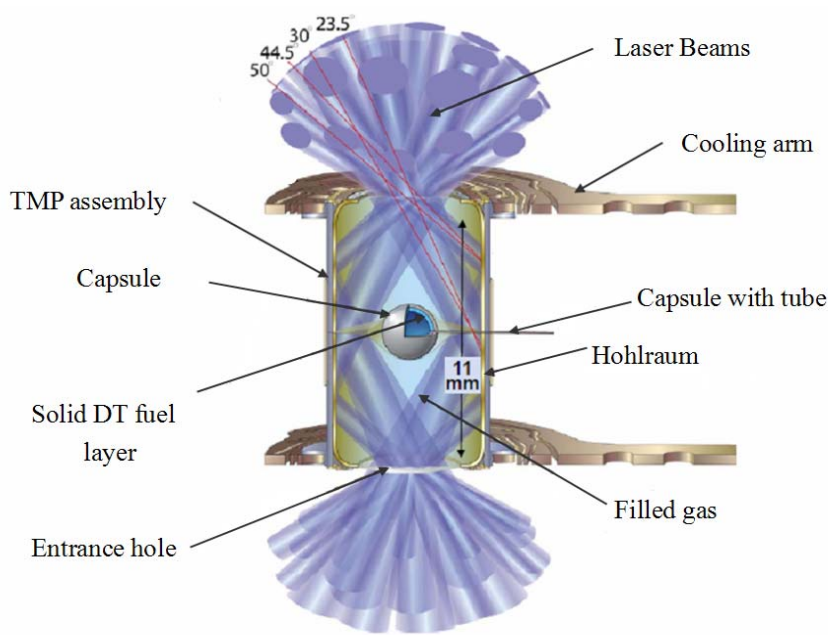

Fig.1. The schematic diagram of the cryogenic target [2]

This paper provides the structure design, the fabrication process and the experimental analyses of the sidewall and mechanical characteristics for cooling arm. The cooling arm in the target plays an important role in thermal conduction and positioning. The thermal conduction requires the cooling arms sufficiently contact with TMP assembly. In order to obtain the maximum contact areas between cooling arm and TMP assembly, the cooling arm with good vertical sidewall is necessary and the cooling arm is designed with 16 clamping arms. As to positioning, the end of cooling arm is elastically clamp TMP assembly to achieve positioning. Because of the small dimension of the cooling arm, conventional machining is very difficult to realize it. Through employing micro-machining technology, optimizing the parameters of ICP etching and designing identical width pattern, the cooling arm with high vertical sidewall was fabricated, and measured by SEM subsequently. Meanwhile, mechanical test shows that this structure of cooling arm can achieve the mechanical characteristics of positioning TMP assembly.

## 2 The design of cooling arm

**2.1 Material selection**

The cooling arm plays an important role in thermal conduction to maintain the required temperature and position the TMP assemble. The required temperature must be stable and isothermal around the capsule. To obtain the required temperature, it requires the material of "cooling-arm" has high thermal conductivity to response the temperature variation quickly. In order to prevent the deformation of the TMP assemble during positioning, the TMP assemble must have a uniform and symmetrical radical clamping force in its circumference. To meet this requirement, the material of the cooling arm with high and isotropic young's modulus was chosen. The electric layers on the top of the cooling arm are connected to heater and temperature sensor. The high electrical resistivity of cooling arm is necessary to avoid the "cooling-arm" surface electric layer interconnection. Therefore, the material of the "cooling-arms" with high thermal conductivity, isotropic young's modulus and high electrical resistivity is needed.

Compared with other materials, the intrinsic silicon has a high thermal conductivity, high young's modulus and electrical resistivity. In terms of silicon, different crystal orientations have different young's modulus [6]. As shown in Fig.2 (a) and Fig.2 (b) for silicon (100) and (110), Young's modulus of silicon (100) varies from 130.2Gpa to 168.9Gpa and that of silicon (110) varies from 130.2 to 187.5Gpa. Silicon (100) and (110) are anisotropic, but in Fig.2 (c) for silicon (111), Young's modulus is transversely isotropic at 168.9Gpa, and Shear modulus is also a constant at $G_v$=57.8Gpa for planes parallel to (111), and $G_p$=57.8Gpa for planes vertical to (111). The Poisson's ratio has a constant value of $\mu_v$=0.262 for planes parallel to (111), and a constant value of $\mu_p$=0.262 for planes vertical to (111) [6]. The yield strength is [S] =7Gpa for the silicon material.

By analyzing the mechanical properties of different crystal orientation silicon, the (100) oriented silicon has anisotropy young's modulus which will result in non-uniform clamping force of the TMP in its circumferential direction. This non-uniform clamping force may cause the deformation of TMP. The crystal orientation (111) has isotropic elastic modulus which will obtain uniform clamping force. Finally, a 500-μm thick (111) oriented intrinsic silicon wafer was used as the substrate to fabricate the cooling arm.

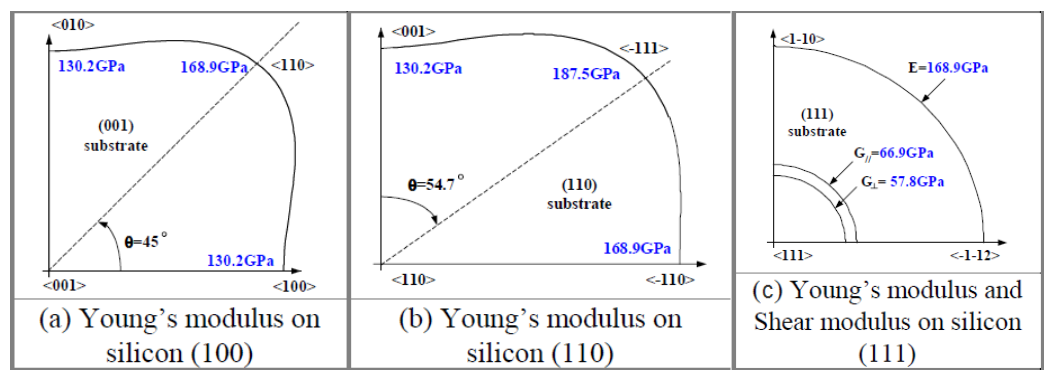

Fig.2. Different mechanical characteristics of different oriented silicon substrates [6]

**2.2 The structural design of cooling arm**

Two pieces of silicon cooling arm are attached to either end of TMP assemble, which are used to cool and position the TMP assembly. The requirements of "cooling-arms" in the target are as follow: (1) Large contact area between cooling arm and TMP assembly is required to ensure maximum heat transfer during cooling the TMP. (2) As the TMP assembly is positioned by cooling arm, arms are designed to elastically clamp the TMP assembly. (3) In order to prevent elastic deformation of TMP assembly due to different clamping force distributed in its circumferential direction, the cooling arm with many arms was required to ensure the symmetry and uniform radial clamping force.

The structure of cooling arm, which contains a positioning ring, was designed to achieve both positioning and heat transfer. The positioning ring, which consists of several clamping arms, is located on one side of cooling arm. As the silicon material was chosen, the good thermal conduction between cooling arm and TMP assembly depends on the contact area themselves. It requires the clamping surfaces of the cooling arm have a uniform curvature with the part of TMP assembly. In order to obtain large contact areas between them, the "cooling-arm" with the largest number of clamping arms is needed. Fig. 3 shows the schematic diagram of the TMP assembly. The difficulty of positioning will be reinforced with the increasing quantity of arms. Considering the maximum number of clamping arms and the difficult of positioning, the "cooling-arm" structure is designed with 16 clamping arms, as shown in Fig.4. In order to achieve the effective positioning of TMP assembly, the clamping surface will yield the displacement of 20μm in its radial direction.

Using the ANSYS finite element analysis, the mechanical characteristics of this cooling arm structure is shown in Fig .5. When there is 20μm displacement in its radial direction of clamping surface, the calculated clamping force is 0.28N and the maximum stress of the clamping arms is 0.209Gpa, which is far less than the silicon material yield strength 7Gpa.

The above analysis shows that this cooling arm structure can meet the mechanical requirement of TMP assembly positioned, and achieve large contact area between TMP assembly and cooling arm.

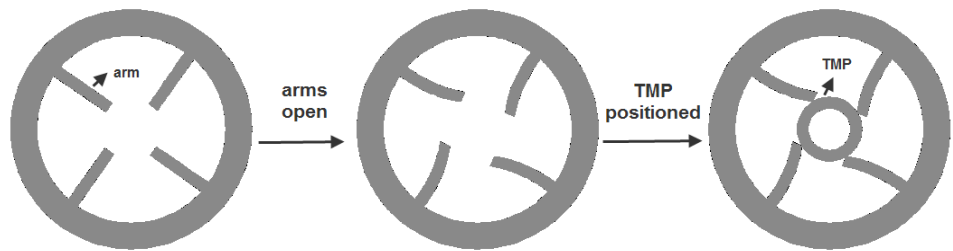

Fig.3. The schematic diagram of the TMP assembly positioned

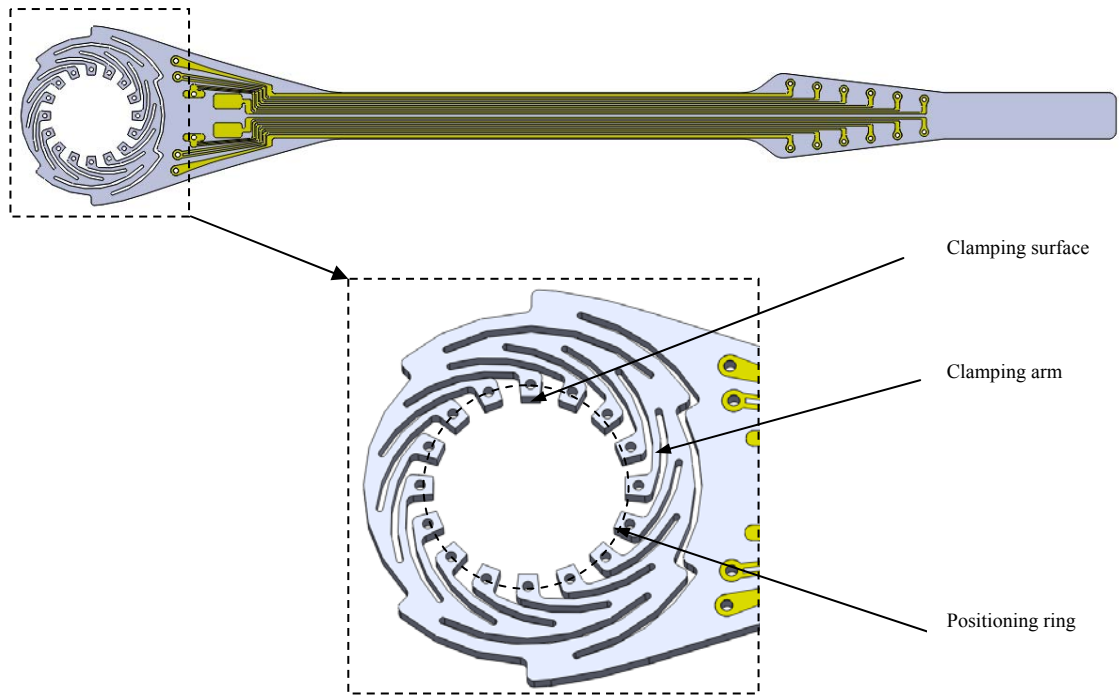

Fig.4. The structure of cooling arm

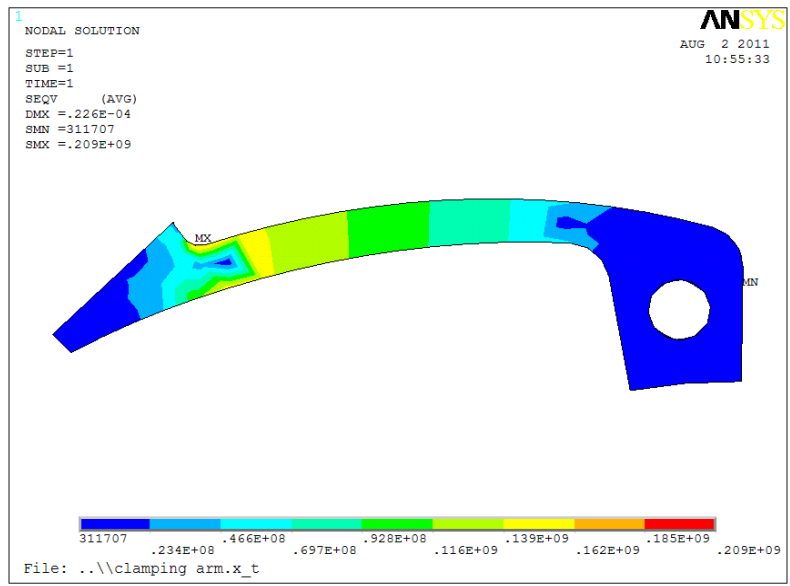

Fig .5. The von Mises Stress graph of clamping arm with the radial displacement of 20μm

## 3 Fabrication process of cooling arm

Fig.6 illustrates the process flow for the cooling arm of silicon structure. A 500μm - thick (111) oriented silicon wafer, double-side polished and oxidized, was used as the substrate with the oxide thickness of 1μm.

In step (1), the wafer is baked and coated an adhesive layer (HMDS) on it. In step (2), the $SiO_2$ layer on the back side of the substrate was removed by using reactive ion etching (RIE). In step (3), approximately 1.5μm of thin positive photoresist was spin-coated on the top of the substrate. The photoresist was prebaked on the hotplate. In step (4), it was exposed by using a contact mask aligner with a UV light source, developed and followed with a post bake. In step (5), the $SiO_2$ layer on the top side of the substrate was patterned by using RIE. The patterned areas were used to place thermal sensor and heater. In step (6), the photoresist layer was removed in acetone and IPA, then repeated step (1). In step (7) and step (8), step (3) and step (4) were repeated respectively. In step (9), the Cr/Au (50/150nm) layers were sputtered on the top of the $SiO_2$ layer as the electric layers which are used to connect sensor and heater. In step (10), the lift-off process was used to form the Cr/Au layers circuit. In step (11), repeated step (1), then approximately 20μm positive photoresist (AZ4620) was spin-coated on the top of the wafer as the deep reactive ion etching (DRIE) mask layer. The resist was prebaked on the hotplate. In step (12), the photoresist was exposed, developed and followed with a post bake. In step (13), the $SiO_2$ layer on the top side of the substrate was patterned for the subsequent DRIE by using RIE. In step (14), through-wafer DRIE was completed. The parameters of the inductively coupled plasma (ICP) etching were optimized to obtain the vertical sidewall. In step (15), the resist was removed in acetone and isopropanol (IPA). The picture of fabricated cooling arm is shown in Fig .7.

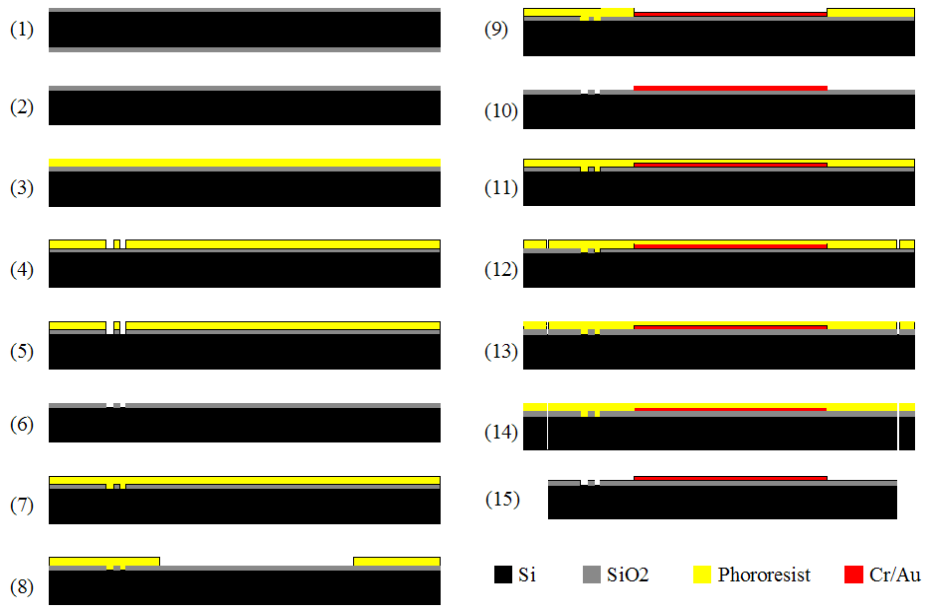

Fig.6. The fabrication-process flow for the cooling arm

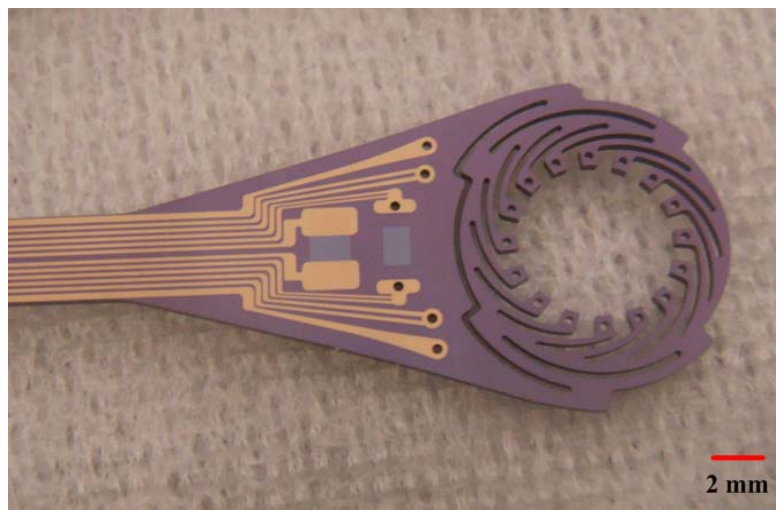

Fig.7. The picture of fabricated cooling arm

# 4 Measurements

**4.1 The sidewall vertical measurement of cooling arm**

As to precise control of heat transfer, the thermal transfer level depends on the contact type between the clamping arms and TMP assembly. In order to ensure the maximum thermal transfer, it required the clamping surface has good sidewall verticality to obtain surface contact with TMP assembly. To obtain the high vertical sidewall, the conditions of ICP etching were optimized as following steps. Firstly, for avoiding the lag effect and obtaining the same sidewall morphology of the cooling arm during ICP etching, mask was patterned as uniform width geometry with all etching occurring in a constant-width trench [7-11]. Secondly, in every etching cycle, the DRIE parameters were optimized by reducing the passivation time to obtain high vertical sidewall [12-15].

The complex structural cooling arm, the uniform width pattern which can result in high vertical sidewall, was fabricated from a 500- thick (111) oriented silicon wafer. The SEM image of complex structural "cooling ring" is showed in Fig.8. This structure has high vertical sidewall and the average angle is in the range of $90\pm0.5°$. The sidewall was measured by SEM and one of the test images was illustrated in Fig.9.

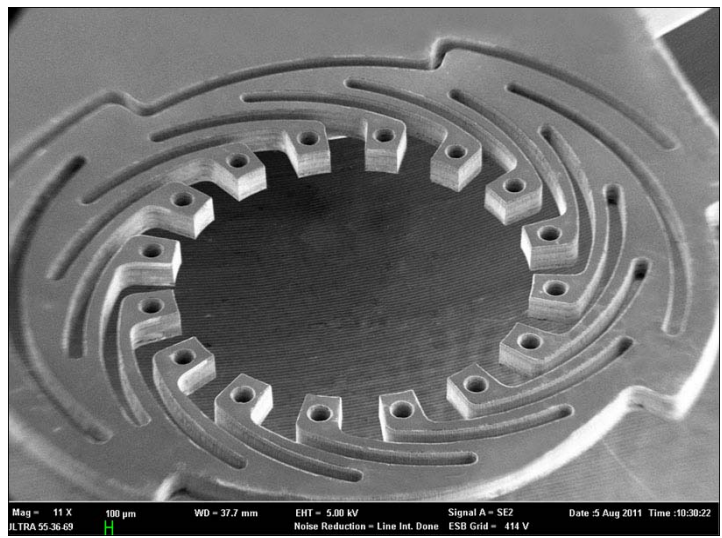

Fig.8. The SEM image of "cooling ring" with high verticality in its sidewall

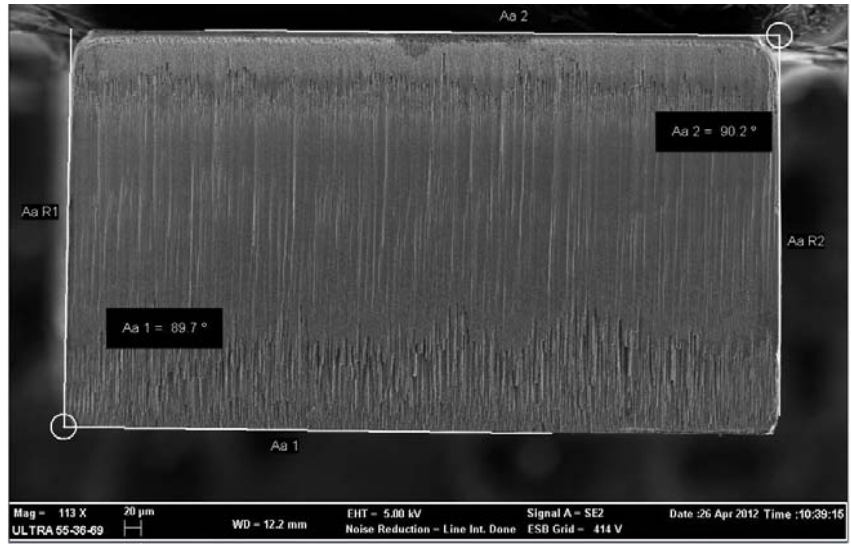

Fig.9. One of the SEM image of sidewall verticality measurement

## 4.2 The mechanical test of clamping arms of cooling arm

For the clamping arms are used to clamp the TMP assembly, it is important to research the mechanical characteristics of the clamping arms. The mechanical measurement equipment （PTR-1101）of clamping arm is showed in Fig.10. One end of cooling arm was fixed and the test probe was placed in front of contact surface when operating the mechanical test equipment. The test probe moved along the radial direction of the contact surface during measuring. Fig.11 is a graph which plots the force versus displacement for the clamping arm. As shown in this graph, the relationship of displacement versus mechanical is approximately linear. In the measurement, when the displacement is 20 μm in the radial direction of contact surface, the measured clamping force is 0.31N, which is approximately consistent with the result of the finite element analysis. The clamping force 0.31 N is far less than the fracture force 0.88N, as shown in Fig. 11. This mechanical test of the clamping arm shows that the characteristics of this "cooling arm" can meet the requirement of the TMP assembly.

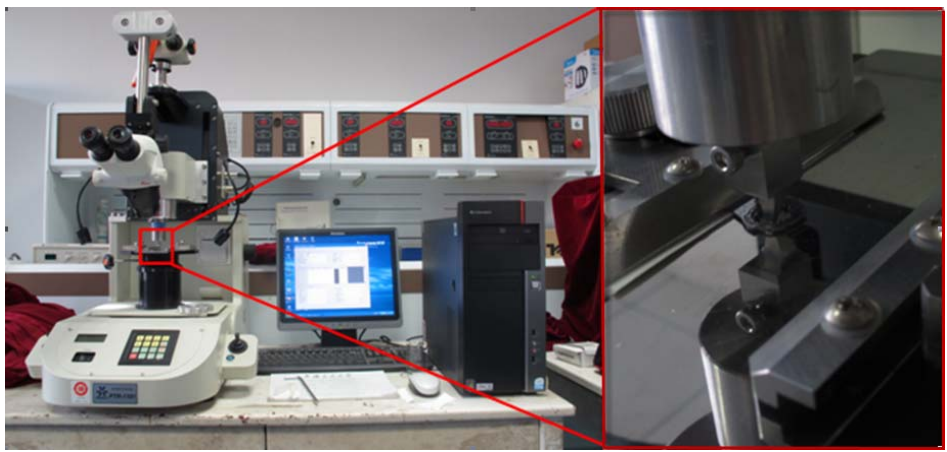

Fig.10. The equipment of the mechanical test

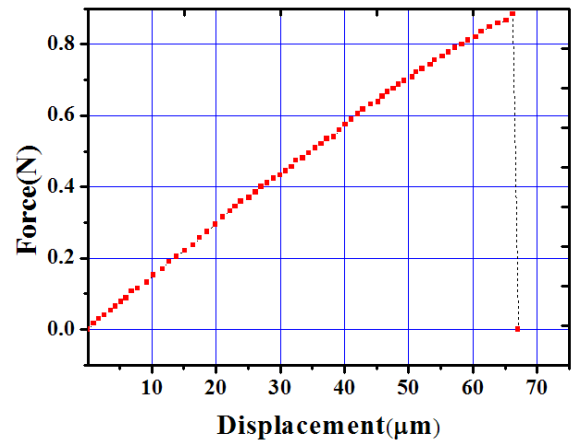

Fig.11. The force versus displacement of the clamping surface

## 5 Conclusion

Through the material selected, structure design and finite element analysis, the cooling arm was fabricated in a silicon (111) oriented wafer. The cooling arm has good vertical sidewall with an approximate angle of 90$^o$ measured by SEM, which meets the requirement of maximum contact between cooling arm and TMP assembly. The mechanical test of clamping arm shows that when the displacement of clamping surface is 20μm in its radial direction, the clamping force is 0.31N which is far less than the fracture force 0.88N. The mechanical characteristics of the clamping arm meet the requirement of positioning TMP assembly in this cooling arm structure.

## Acknowledgments

This work is partly supported by the National Natural Science Foundation of China (No. 61076107)，the Science and Technology Department of Shanghai (No. 11DZ2290203，11JC1405700)， Program for New Century Excellent Talents in University (2009).The authors are also grateful to the colleagues for their essential contribution to this work.